\documentclass[prl,showpacs,twocolumn]{revtex4}
\usepackage{tabularx}
\usepackage{amsmath}
\usepackage{amssymb}
\usepackage{graphicx}
\usepackage{dcolumn}
\usepackage{bm}
\usepackage{ulem}

\begin{document}

\title{New Route to Observable Fulde-Ferrell-Larkin-Ovchinnikov Phases in 3D
Spin-Orbit Coupled Degenerate Fermi Gases}
\author{Zhen Zheng$^{1}$}
\author{Ming Gong$^{2}$}
\thanks{Email: skylark.gong@gmail.com}
\author{Xubo Zou$^{1}$}
\thanks{Email: xbz@ustc.edu.cn}
\author{Chuanwei Zhang$^{2}$}
\thanks{Email: chuanwei.zhang@utdallas.edu}
\author{Guangcan Guo$^{1}$}

\begin{abstract}
The Fulde-Ferrell-Larkin-Ovchinnikov (FFLO) phase was first predicted in 2D
superconductors about 50 years ago, but so far unambiguous experimental
evidences are still lacked. The recently experimentally realized
spin-imbalanced Fermi gases may potentially unveil this elusive state, but
require very stringent experimental conditions. In this Letter, we show that
FFLO phases may be observed even in a 3D degenerate Fermi gas with
spin-orbit coupling and in-plane Zeeman field. The FFLO phase is driven by
the interplay between asymmetry of Fermi surface and superfluid order,
instead of the interplay between magnetic and superconducting order in solid
materials. The predicted FFLO phase exists in a giant parameter region,
possesses a stable long-range superfluid order due to the 3D geometry, and
can be observed with experimentally already achieved temperature ($T\sim
0.05E_{F}$), thus opens a new fascinating avenue for exploring FFLO physics.
\end{abstract}

\affiliation{$^{1}$Key Laboratory of Quantum Information, University of Science and
Technology of China, Hefei, Anhui, 230026, People's Republic of China \\
$^{2}$Department of Physics, the University of Texas at Dallas, Richardson,
TX, 75080 USA}
\maketitle


The Fulde-Ferrell-Larkin-Ovchinnikov (FFLO) phase, characterized by Cooper
pairs with finite total momentum and spatially modulated order parameters,
was predicted to exist in certain region of 2D superconductors in high
Zeeman fields \cite{Ferrell64,Larkin64,Larkin65}. This fascinating state
arises from the interplay between magnetic and superconducting order, and
now is a central concept for understanding many exotic phenomena in
different physics branches \cite{Buzdin,
Croitoru,Yuji,Gloos,Bianchi,Singleton, Lortz,Casalbuoni,Alford}. Despite
tremendous experimental and theoretical efforts in the past five decades,
there is still no unambiguous experimental evidence for FFLO states \cite%
{Casalbuoni}. The experimental difficulty may arise from several different
aspects, such as the depairing of Cooper pairs due to orbital or Pauli
effects in strong magnetic fields and unavoidable disorder effects in solid
state materials.

The recent experimental realization of spin-imbalanced Fermi gases \cite%
{Zwierlein05,Zwierlein06,Hulet1,Hulet2,Shin} provides a new excellent
platform for exploring FFLO physics. In Fermi gases, the effective Zeeman
field is generated through the population imbalance between two spins,
therefore the orbital effects (e.g., vortices induced by the magnetic field)
are absent even in 3D. The Fermi gases are also free of disorder and all
experimental parameters are highly controllable. These advantages have
sparked tremendous recent interest in exploring FFLO physics in
spin-imbalanced Fermi gases \cite{Hu, He06,
Parish,Melo,Bulgas,Sheehy,Koponen,Liu, Yasuharu, Devreese1, Devreese2}.
However, the FFLO phase only exists in a narrow parameter regime in 3D due
to the Pauli depairing effect \cite{Hu,Bulgas,Sheehy}. Furthermore, the free
energy difference between the FFLO state and the BCS superfluid is extremely
small. As a result, only the transition from the BCS superfluid to the
normal gas \cite{Zwierlein05,Zwierlein06,Hulet1} has been observed in
experiments in 3D spin-imbalanced Fermi gases. Current experimental and
theoretical efforts on the FFLO state have focused on low dimensions (1D or
2D) \cite{Hulet3,Orso,Hu2,Parish2,Pu}, where quantum and thermal (at finite
temperature) fluctuations may become crucial and the physics is much more
complicated \cite{BKT1,BKT2,BKT3}.

In this Letter we show that a large and stable parameter region for FFLO
states can be realized even in a 3D degenerate Fermi gas by including two
experimentally already developed \cite{Spielman,Jing,Zwierlein12} elements:
spin-orbit (SO) coupling and an in-plane Zeeman field. Recently the BCS-BEC
crossover physics of SO coupled Fermi gases with perpendicular Zeeman fields
has been intensively investigated with the goal of realizing topological
superfluids \cite{CWZhang,Gong11,JZhou,Iskin} and the associated Majorana
fermions \cite{Gong12,LiuXJ,Iskin2}. However, regular BCS superfluids,
instead of FFLO states, are energetically preferred for perpendicular Zeeman
field because of the centrally symmetric Fermi surface. We show that this
issue can be resolved by using an in-plane Zeeman field, which, together
with the SO coupling, yields an asymmetric Fermi surface so that the FFLO
state can emerge naturally. We emphasize that here the FFLO phase is driven
by the asymmetry of the Fermi surface, instead of spin imbalance in previous
study in 2D superconductors and Fermi superfluids. More importantly, we find
that the energy difference between the FFLO ground state and the possible
BCS superfluid excited state is dramatically increased (to $\sim 0.04E_{F}$
per particle), therefore the FFLO state is experimentally more accessible
with the realistic temperature in 3D ($T\sim 0.05E_{F}$). Finally, because
of the 3D geometry, the quantum and thermal fluctuations that play major
roles in 1D and 2D are strongly suppressed \cite{BKT1,BKT2,BKT3}, which
greatly simplifies the FFLO physics. Finally, we argue that our system has
no direct solid state analogy and the new route represents a more efficient
way to create and observe FFLO phases.

\textit{Thermodynamical potential:} Consider a 3D degenerate Fermi gas in
the presence of a Rashba type of SO coupling and an in-plane Zeeman field.
The corresponding partition function of the system can be expressed as $Z=%
\text{Tr}e^{-\beta (H-\mu N)}=\int \mathcal{D}\psi e^{-S}$, with the action $%
S=\int \psi ^{\dagger }(\partial _{\tau }+H_{0})\psi +g\psi _{\uparrow
}^{\dagger }\psi _{\downarrow }^{\dagger }\psi _{\downarrow }\psi _{\uparrow
}$. Here $\int =\int_{0}^{\beta }d\tau d^{3}\mathbf{r}$, $\psi ^{\dagger
}=(\psi _{\uparrow }^{\dagger },\psi _{\downarrow }^{\dagger })$, $H_{0}={%
\frac{\mathbf{p}^{2}}{2m}}-\mu -h\sigma _{x}+\alpha (p_{x}\sigma
_{y}-p_{y}\sigma _{x})$, $m$ is the mass of the atom, $\mu $ is the chemical
potential, $g$ is the $s$-wave interaction strength, $\alpha $ is the Rashba
SO coupling strength, and $h$ is the in-plane (same as the SO coupling)
Zeeman field. In experiments, the SO coupling and the in-plane Zeeman field
can be realized using the tripod scheme where three Raman lasers couple
three hyperfine ground states with a common excited state \cite%
{Tudor,Dalibard,YP,note}. Note that an in-plane Zeeman field is generated
naturally using three Raman lasers in the tripod scheme \cite%
{Tudor,Dalibard,YP}, while a perpendicular Zeeman field requires additional
lasers \cite{Zhang2} (thus more difficult in experiments).

\begin{figure}[t]
\includegraphics[width=3.2in]{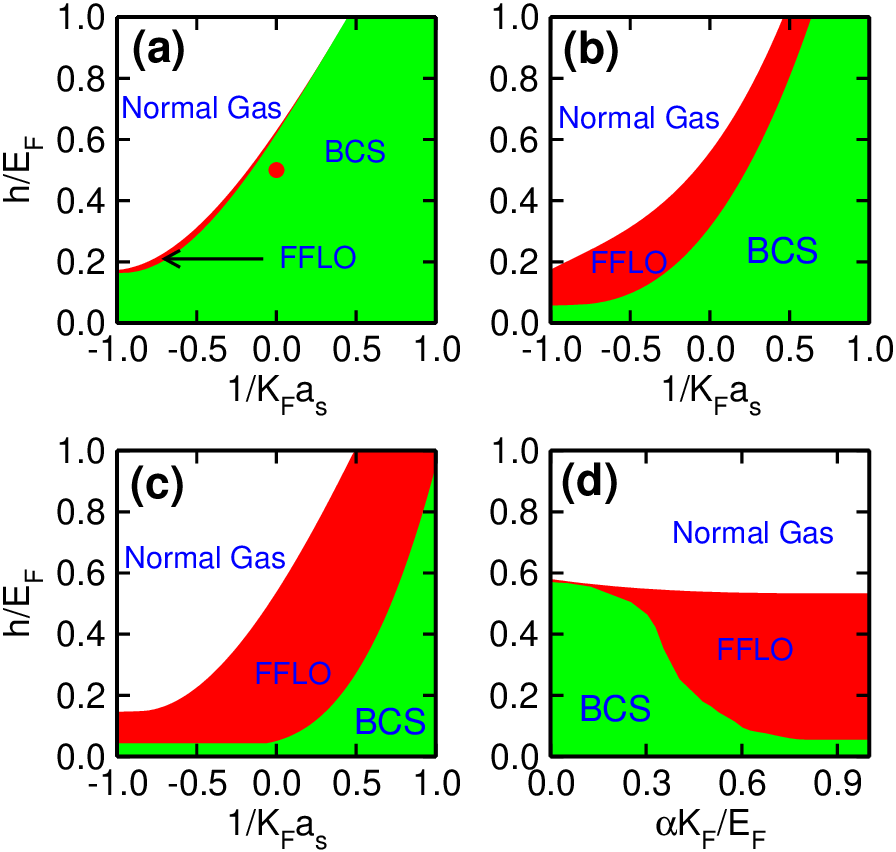} \centering
\caption{BCS-BEC crossover phase diagrams in the presence of SO coupling and
in-plane Zeeman field. (a) Without SO coupling. The circle symbol represents
the data from the quantum Monte Carlo calculation \protect\cite{Carlson}.
(b) and (c) With $\protect\alpha K_{F}=0.5E_{F}$ and $\protect\alpha %
K_{F}=1.0E_{F}$. (d) In the unitary regime.}
\label{fig-phase}
\end{figure}

In the FFLO state the Cooper pairs have finite total momentum, \textit{i.e.}%
, $\Delta (\mathbf{r})=\langle \psi _{\downarrow }\psi _{\uparrow }\rangle
=\Delta e^{i\mathbf{Q}\cdot \mathbf{r}}$, where $\mathbf{Q}$ is the FFLO
vector. We adopt a spatial uniform order parameter $\Delta $ in our
calculation through a transformation of the field $\psi \rightarrow \psi e^{i%
\mathbf{Q}\cdot \mathbf{r}/2}$, yielding a new Hamiltonian $e^{i\mathbf{Q}%
\cdot \mathbf{r}/2}H_{0}(\mathbf{p})e^{i\mathbf{Q}\cdot \mathbf{r}/2}=H_{0}(%
\mathbf{p}+\mathbf{Q}/2)=\bar{H}_{0}$, hence%
\begin{equation}
S=\int \psi ^{\dagger }(\partial _{\tau }+\bar{H}_{0})\psi -|\Delta
|^{2}/g+\Delta \psi _{\uparrow }^{\dagger }\psi _{\downarrow }^{\dagger
}+\Delta ^{\dagger }\psi _{\downarrow }\psi _{\uparrow }  \label{eq-S1}
\end{equation}%
in the new field basis. Integrating out the Fermi field, we obtain $Z=\int
\mathcal{D}\Delta \exp (-S_{\text{eff}})$, with the effective action
\begin{equation}
{\frac{S_{\text{eff}}}{\beta }}=-{\frac{|\Delta |^{2}}{g}}-\sum_{\lambda ,%
\mathbf{k},i\omega _{n}}{\frac{\ln \beta (i\omega _{n}-E_{\lambda })}{2\beta
}}+\sum_{\mathbf{k},\sigma }{\frac{\xi _{{\frac{\mathbf{Q}}{2}}-\mathbf{k}%
,\sigma }}{2}},  \label{eq-eff}
\end{equation}%
where $\xi _{{\frac{\mathbf{Q}}{2}}-\mathbf{k},\sigma }=({\frac{\mathbf{Q}}{2%
}}-\mathbf{k})^{2}/2m-\mu $, $E_{\lambda }$ ($\lambda =1$, 2, 3, 4) are the
eigenstates of the effective Hamiltonian (under the basis ($\psi _{\mathbf{Q}%
/2+\mathbf{p},\uparrow }$, $\psi _{\mathbf{Q}/2+\mathbf{p},\downarrow }$, $%
\psi _{\mathbf{Q}/2-\mathbf{p},\downarrow }^{\dagger }$, $-\psi _{\mathbf{Q}%
/2-\mathbf{p},\uparrow }^{\dagger })^{T}$)
\begin{equation}
H_{\text{eff}}(\mathbf{k},\mathbf{Q})=%
\begin{pmatrix}
H_{0}({\frac{\mathbf{Q}}{2}}+\mathbf{k}) & \Delta \\
\Delta ^{\dagger } & -\sigma _{y}H_{0}^{\ast }({\frac{\mathbf{Q}}{2}}-%
\mathbf{k})\sigma _{y}%
\end{pmatrix}%
.
\end{equation}%
In Eq. (\ref{eq-eff}) the bare interaction strength $g$ should be
regularized in terms of the $s$-wave scattering length $a_{s}$ \cite{Gong11,
Iskin}, ${\frac{1}{4\pi \hbar a_{s}}}={\frac{1}{g}}+\sum_{\mathbf{k}}{\frac{1%
}{2\epsilon _{\mathbf{k}}}}$, where $\epsilon _{\mathbf{k}}={\frac{k^{2}}{2m}%
}$.

The ground state phase diagram of the system (i.e., $\Delta $, $\mu $, $%
\mathbf{Q}$) is determined by the saddle point of the thermodynamical
potential ${\frac{\partial \Omega }{\partial \Delta }}=0$ and ${\frac{%
\partial \Omega }{\partial \mathbf{Q}}}=0$, as well as the atom number
conservation $n=\sum_{\sigma =\uparrow ,\downarrow }n_{\sigma }=-{\frac{%
\partial \Omega }{\partial \mu }}$, where $\Omega =S_{\text{eff}}/\beta $.
The energy unit is chosen as the Fermi energy $E_{F}$ for an non-interacting
gas without SO coupling and Zeeman field. The length unit is $K_{F}^{-1}$.
We restrict to $T=0$ throughout this work. Generally the vector $\mathbf{Q}$
has three different components, and the total five unknown parameters put a
great burden for numerically solving the above equations self-consistently
because the landscape of $\Omega $ is an extremely complex function of these
parameters whose global minimum (instead of a local minimum) is hard to
find. For the $x$-axis Zeeman field and the Rashba-type SO coupling the
deformation of the Fermi surface is along the $y$-axis, therefore the FFLO
vector is expected to be along the $y$ axis, \textit{i.e.}, $\mathbf{Q}%
=(0,Q,0)$. We have numerically confirmed that there is no large FFLO region
when $\mathbf{Q}$ is along the $x$ and $z$ directions. There are three
possible phases in this system: BCS superfluid ($\Delta \neq 0$, $Q=0$) (we
still use BCS for convenience although we really consider the BCS-BEC
crossover physics), FFLO ($\Delta \neq 0$, $Q\neq 0$), and normal gas ($%
\Delta =0$ and $Q=0$). In the FFLO phase, we also calculate the energy
difference between the FFLO ground state and the possible BCS superfluid
excited state (by enforcing $Q=0$) to check the stability of the FFLO state
against the finite temperature effect.

\begin{figure}[t]
\centering
\includegraphics[width=3.0in]{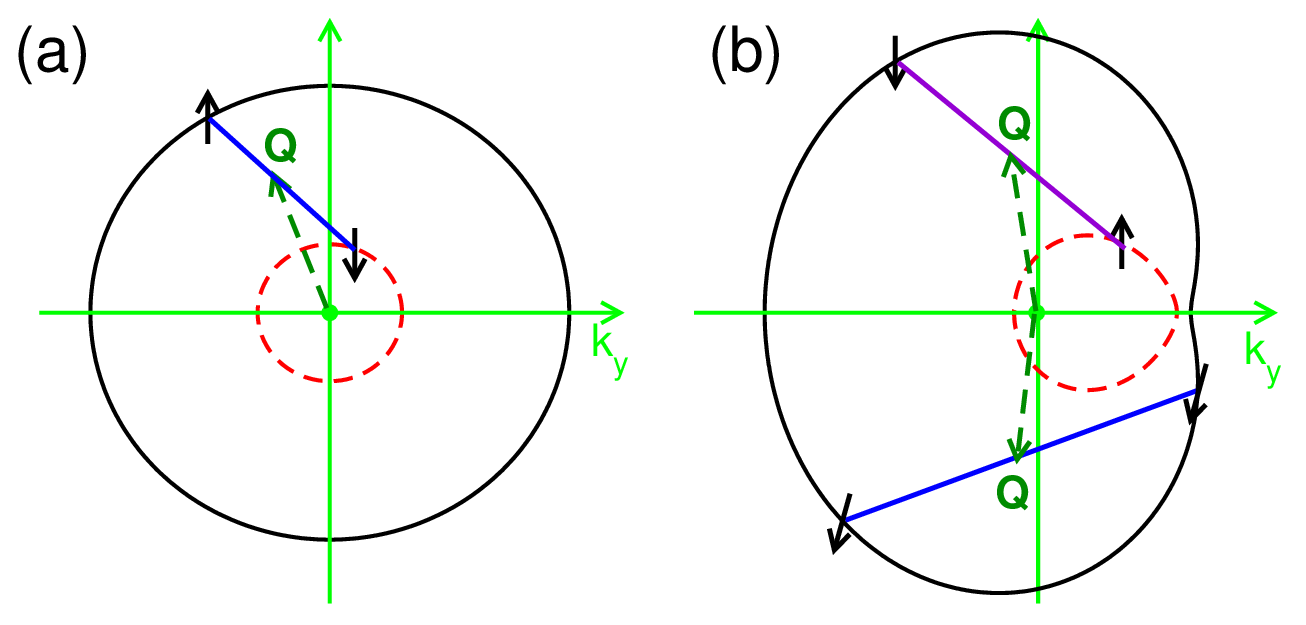}
\caption{Illustration of the physical mechanism of the FFLO state in the
presence of an in-plane Zeeman field and SO coupling. Solid and dashed
contours are two Fermi surfaces. The solid arrows are the pseudospins. The
solid line connecting two pseudospins represents the Cooper pair with total
momentum $\mathbf{Q}$ (direction is shown by the dashed arrow). (a) Without
SO coupling, the Fermi surfaces are two concentric spheres. (b) With SO
coupling, the Fermi surfaces are anisotropic along the $k_{y}$ axis due to
the Rashba SO coupling and the $x$-axis Zeeman field.}
\label{fig-idea}
\end{figure}

\textit{Phase diagram and mechanism for FFLO phase}: In Fig. \ref{fig-phase}%
, we plot the phase diagrams of the Fermi gas with respect to $h$, $%
1/K_{F}a_{s}$, and $\alpha K_{F}$. Without SO coupling (Fig. \ref{fig-phase}%
a), our numerical result agrees well with that in previous literatures using
the mean-field approximation \cite{Hu} or quantum Monte Carlo \cite{Carlson}%
. We see the FFLO phase exists only within an extremely small regime in the
phase diagram. Furthermore, the energy difference per particle between the
FFLO state and the possible BCS superfluid state (obtained by enforcing $Q=0$%
) is extremely small (see Fig. \ref{fig-crossover}d), therefore the Fermi
gas may not relax to the FFLO state considering the realistic temperature in
experiments \cite{Zwierlein05}, even if the FFLO state is the true ground
state. With increasing SO strength, the parameter region of the FFLO phase
is greatly enlarged. In Fig. \ref{fig-phase}d, we see that the critical
Zeeman field for the transition between the BCS superfluid and FFLO phase is
greatly reduced, but always larger than zero because of the required
time-reversal symmetry breaking for the FFLO phase.

\begin{figure}[tbp]
\centering
\includegraphics[width=3.3in]{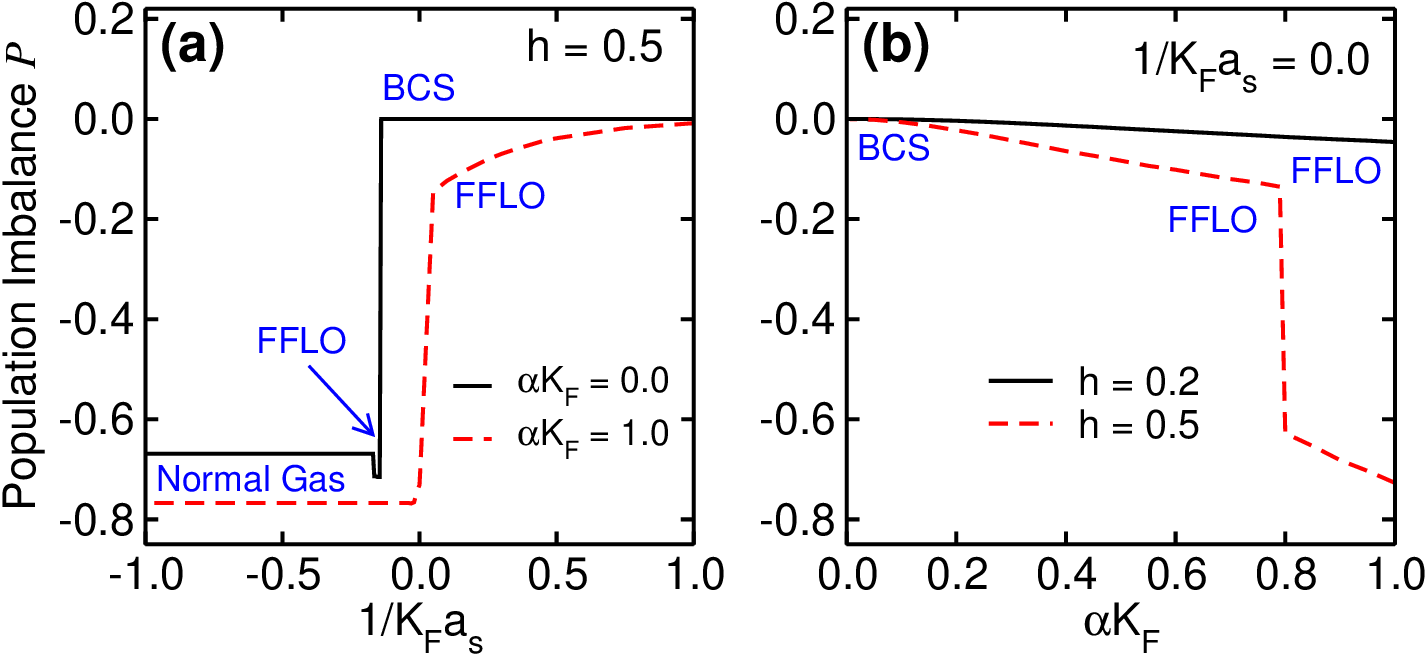}
\caption{Population imbalance $P=\protect\delta n/n$ as a function of the
scattering interaction (a) and the SO coupling strength (b).}
\label{fig-sigmax}
\end{figure}

The enlarged parameter region for the FFLO state in Fig. \ref{fig-phase} can
be understood from the change of the shape of the Fermi surface due to the
SO coupling and the in-plane Zeeman field. Without the SO coupling, the
Zeeman field (no matter which direction) yields two concentric spheres (Fig. %
\ref{fig-idea}a) of the Fermi surface, and only singlet pairing between
different pseudospins (i.e., two eigenstates of $H_{0}$) is allowed due to
the SU(2) symmetry of the Hamiltonian. With increasing Zeeman fields, the
Fermi surface mismatch increases the energy cost of the BCS superfluid. In a
strong Zeeman field the superfluid has to break the spatial symmetry to
lower the accumulated energy, therefore the FFLO state emerges, but only
survives in a small parameter region due to the Pauli depairing effect. Such
depairing effect in strong Zeeman fields can be circumvented using the SO
coupling, which allows both singlet and triplet pairings \cite%
{Gong11,Hu3,Zhai} (the later is insensitive to the depairing effect) because
the pseudospin state is a spin mixed state with strong momentum dependence
\cite{CWZhang}. However, if a perpendicular Zeeman field is applied, the
regular BCS superfluid in the same SO band (i.e., triplet pairings)
dominates because of the symmetric Fermi surface, and the parameter region
for the FFLO state indeed shrinks comparing with that with only Zeeman
fields, as found in our numerical simulation. In contrast, in the presence
of SO coupling and an in-plane Zeeman field, the Fermi surfaces become
anisotropic and the center of the Fermi surface is also shifted accordingly
(Fig. \ref{fig-idea}b). Therefore the regular BCS superfluid, which is
preferred for a symmetric Fermi surface, is greatly suppressed, and the FFLO
state becomes energetically favorable in a much wider parameter region, as
observed in Fig. \ref{fig-phase}. Note that without SO coupling (Fig. \ref%
{fig-idea}a), the system has the rotation symmetry, therefore $Q$ can be
along any direction. The SO coupling breaks the rotation symmetry, and
forces $Q$ to the direction of the asymmetric axis (thus $Q$ is unique).

The FFLO phase is induced by the interplay between asymmetry of Fermi
surface and superfluid order, instead of the interplay between magnetism%
\textbf{\ }and superconducting order in solid materials. In our model the
only spin polarization is along the $\sigma _{x}$ axis (i.e., $\langle
\sigma _{z}\rangle =0$, $\langle \sigma _{y}\rangle =0$), thus we define the
the population imbalance as $P=\delta n/n$ with $\delta n=\langle \sigma
_{x}\rangle $. In Fig. \ref{fig-sigmax}, we plot $P$ with respect to $%
1/K_{F}a_{s}$ and $\alpha K_{F}$. Without SO coupling, the BCS superfluid
breaks down at $P\sim 0.669$, in consistent with previous results \cite%
{Hu,Zwierlein05}. When the SO coupling is applied, the FFLO phase can emerge
with a much smaller population imbalance ($P\sim $ 0.1 - 0.2).\textbf{\ }From Fig. %
\ref{fig-sigmax}, we see that the SO coupling generally enhances the
population imbalance in the normal phase, however the emergence of the FFLO
does not ocuur in this large population imbalance region. Therefore the FFLO
phase cannot originate from the interplay between magnetism and superconducting
order which is the major driving force for tranditional FFLO superfluid in 
the spin-polarized Fermi gas (withoug SO coupling) and 2D solid state materials.

\textit{Stability and measurement of of FFLO phase}: To characterize the
FFLO state, in Figs. \ref{fig-crossover}, we plot the chemical potential $%
\mu $ and the order parameter $\Delta $ in the BCS-BEC crossover. For
comparison, we also plot $\mu $ and $\Delta $ for the possible BCS
superfluid state (by enforcing $Q=0$). In the weak BCS limit $\Delta $ is
exponentially small, therefore a small population imbalance can destroy the
superfluid \cite{Zwierlein06}. In the BEC side, the fermions form tightly
bound molecules and the influence of Zeeman field and SO coupling is
negligible. Therefore the only relevant parameter regime for the observation
of FFLO states should be near the unitary regime. In the FFLO regime, $%
\Delta $ for the FFLO state is smaller than that for the assumed BCS
superfluid to reduce the FFLO energy. In Fig. \ref{fig-crossover}c, we plot $%
Q$ versus the scattering interaction, which also confirms that the SO
coupling can greatly increase the parameter region for the FFLO phase.

An experimentally observable FFLO state requires a large energy difference
between the FFLO ground state and the possible BCS superfluid excited state
so that the FFLO state can survive at finite temperature. In Fig. \ref%
{fig-crossover}d, we plot the free energy difference between FFLO state and
the BCS superfluid per particle, $\delta F=\left( F_{\text{FFLO}}-F_{\text{%
BCS}}\right) /nE_{F}$, with $F=\Omega +\mu n$. The stability of the FFLO
state has not been emphasized in previous literatures \cite{Koponen,
Yasuharu, Devreese1, Devreese2, He06, Bulgas,Sheehy}. For FFLO states
without SO coupling we find $\delta F$ $\sim 10^{-4}E_{F}$, which is much
smaller than the experimental coldest temperature ($T\sim 0.05E_{F}$) \cite%
{Jin,Zwierlein05}. Therefore the FFLO state cannot be observed even the
exact parameter region has been reached. While with the SO coupling and
in-plane Zeeman field, the energy difference is greatly enhanced to $\sim
0.04E_{F}$, which makes the FFLO state accessible with realistic
experimental temperature. Such a large energy difference is another major
advantage of our scheme over previous Zeeman field \cite{Hu, He06,
Parish,Melo,Bulgas,Sheehy} or optical lattice \cite{Koponen, Yasuharu,
Devreese1, Devreese2} schemes.

\begin{figure}[t]
\centering
\includegraphics[width=3.in]{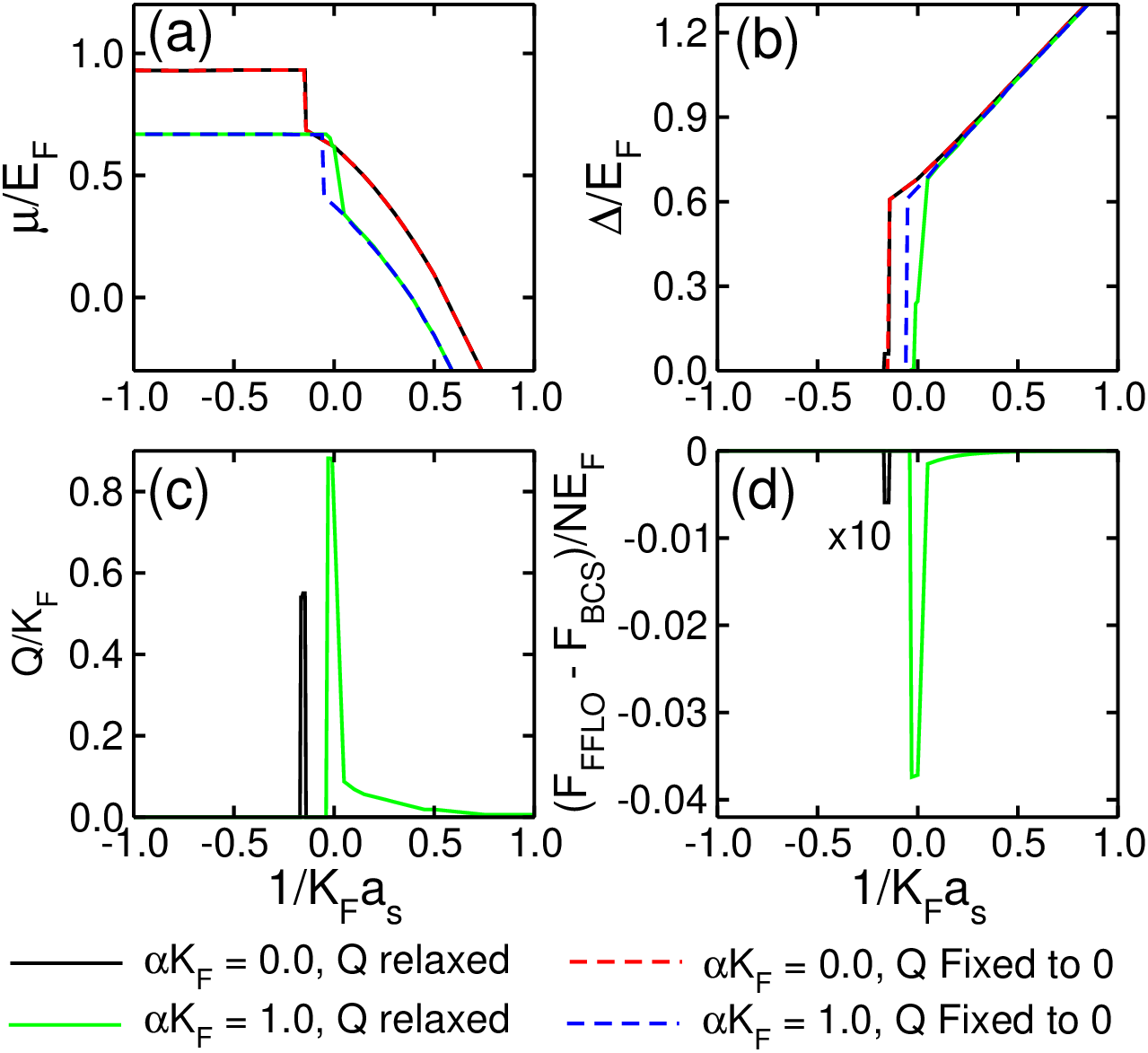}
\caption{BEC-BCS crossover in the presence of SO coupling and an in-plane
Zeeman field. $h=0.5E_{F}$ and $\protect\alpha K_{F}=0.0$ and 1.0$E_{F}$. In
(a) and (b), the solid lines are obtained by minimizing the total free
energy with respect to $\Delta $, $\protect\mu $ and $\mathbf{Q}$, while the
dashed lines are obtained by enforcing $Q=0$ (thus no FFLO states). (c) Plot
of $Q$ as a function of the scattering interaction. (d) The free energy ($F
= \Omega + \protect\mu n$) difference between the FFLO state and the
possible BCS superfluid. }
\label{fig-crossover}
\end{figure}

So far we only consider the FFLO state using a simplified pairing $\Delta (%
\mathbf{r})=\Delta e^{i\mathbf{Q}\cdot \mathbf{r}}$, while the true pairing
of the FFLO state may be different. Because the FFLO state depends strongly
on the nesting of the Fermi surface, the order parameter may be composed of
\ multiple vectors \cite{Yuji,Bulaevskii}, \textit{i.e.}, $\Delta (\mathbf{r}%
)=\Delta (\mathbf{Q}_{1},\mathbf{Q}_{2},\cdots )$ (e.g., the LO state with $%
\mathbf{Q} $ and $-\mathbf{Q}$), whose stability depends strongly on the
detailed structure of the Fermi surface and thus cannot been ruled out \cite%
{Bulaevskii}. However, the main conclusion of our work, the large parameter
region and the stable FFLO state induced by SO coupling and in-plane Zeeman
field, is in intact even for very complex pairings because different choices
of the order parameter are mainly used to further reduce the total energy of
the FFLO state (thus further enhance our results).

The FFLO wavevector $\mathbf{Q}$ may be measured directly using the
time-of-flight images \cite{Koponen, Stein}, where momentum distribution
shows a peak at $\mathbf{r} = \hbar \mathbf{Q}t/m$. Because $\mathbf{Q}$ is
unique in our system, repeated measurements can be used to determine $%
\mathbf{Q}$ precisely. The superfluidity of the FFLO states can be
demonstrated through the rotation of the system, where the generated
vortices provides unambiguous signature of superfluidity \cite{Zwierlein052}%
. Near the boundary of different phases the vortices may be unstable due to
strong damping effects \cite{Zwierlein05}, however in the middle of the FFLO
phase (only possible with a large parameter region for the FFLO state), we
expect the damping effect to be small, similar as that in the BCS superfluid
state.

\textit{Comparision to solid state systems}: We emphasize that our system
has no direct solid state analogy although we note that FFLO states were
also studied recently in 2D spin-orbit coupled superconductors with in-plane
magnetic fields \cite{2DSC1,2DSC2,2DSC3}. Our scheme is different from these
2D superconductors in the following aspects:

(I) Different driving mechanism: in 2D superconductors, FFLO phases are
mainly induced by strong magnetic field and the role of SO coupling is to
enhance the second critical magnetic field $H_{c2}$ between superconducting and
normal states \cite{2DSC1,2DSC2}. While in our 3D Fermi gases, FFLO phases
are induced by the asymmetric Fermi surface, and are present even with a
weak Zeeman field (see Fig. \ref{fig-phase}d) and a small population
imbalance (see Fig. \ref{fig-sigmax}). (II) 3D vs 2D: It is well known that
in 2D and at finite temperature the mean field theory does not work and
there is no long range superconducting order (including FFLO) due to phase
fluctuations \cite{BKT1,BKT2,BKT3}. In contrast, the mean-field theory works
well, at least qualitatively, in 3D degenerate Fermi gases, which can have
long-range FFLO order at finite temperature. (III) BCS-BEC crossover vs BCS
limit: Our study focuses on the BCS-BEC crossover physics (see Figs. 1, 3,
and 4), in contrast with the BCS limit in 2D superconductors \cite%
{2DSC1,2DSC2,2DSC3}; (IV) Different experimental concerns: cold atomic gases
are disorder free and FFLO phases can be observed directly in time-of-flight
images; while in 2D superconductors disorder effects are important \cite%
{2DSC2} and FFLO states can only be observed indirectly.

In summary, we show that the combination of SO coupling and in-plane Zeeman
field can lead to a large and stable parameter region for the experimentally
long-sought FFLO state even for 3D degenerate Fermi gases. Considering the
recent experimental progress on the generation of the SO coupling in Bose
and Fermi gases, our work provides a new exciting research direction for the
study of SO coupled Fermi gases as well as the FFLO physics, which is
essential for the understanding of important phenomena in many branches of
physics, ranging from solid state superconductors to astrophysics.

\textit{Acknowledgement:} Z.Z, X.Z, and G.G are supported by the National
973 Fundamental Research Program (Grant No. 2011cba00200), the National
Natural Science Foundation of China (Grant No. 11074244). M.G and C.Z are
supported by ARO (W911NF-12-1-0334), DARPA-YFA (N66001-10-1-4025), AFOSR
(FA9550-11-1-0313), and NSF-PHY (1104546).


\begin{thebibliography}{99}
\bibitem{note} Without SO coupling, the Zeeman field can be applied through
adjusting the population imbalance between two spin states. While in the
presence of SO coupling, the number of atoms at each spin channel is not
conserved anymore, and the population imbalance can be tuned only through
the external Raman lasers (Zeeman field).
\end{thebibliography}
\end{document}